\documentclass[manuscript, screen]{acmart}
\renewcommand\footnotetextcopyrightpermission[1]{} 
\AtBeginDocument{%
  }

\usepackage{natbib}

\begin{document}

\title{Data Ecofeminism}

\author{Ana Valdivia}
\affiliation{
\institution{Oxford Internet Institute (University of Oxford)
and Institute of Advanced Studies (University College London)}
\country{UK}}
\email{ana.valdivia@oii.ox.ac.uk}

\renewcommand{\shortauthors}{}

\begin{abstract}
Generative Artificial Intelligence (GenAI) is driving significant environmental impacts. The rapid development and deployment of increasingly larger algorithmic models capable of analysing vast amounts of data are contributing to rising carbon emissions, water withdrawal, and waste generation. Generative models often consume substantially more energy than traditional models, with major tech firms increasingly turning to nuclear power to sustain these systems---an approach that could have profound environmental consequences.

This paper introduces seven data ecofeminist principles delineating a pathway for developing technological alternatives of eco-societal transformations within the AI research context. Rooted in data feminism and ecofeminist frameworks, which interrogate about the historical and social construction of epistemologies underlying the hegemonic development of science and technology that disrupt communities and nature, these principles emphasise the integration of social and environmental justice within a critical AI agenda. The paper calls for an urgent reassessment of the GenAI innovation race, advocating for ecofeminist algorithmic and infrastructural projects that prioritise and respect life, the people, and the planet.

\end{abstract}





\maketitle

\section{Introduction}

In 1979, twelve women came together to organise a meeting focused on the relationship between women and ecology, prompted by one of the worst nuclear accidents in history---the partial meltdown at the Three Mile Island nuclear plant in Pennsylvania, United States (US). The incident resulted in the release of radioactive gases from Unit 2, one of the plant's reactors. In response, local authorities evacuated 150,000 residents, including pregnant women and children, and ordered the closure of schools. Motivated by these events, these women formed the group called \textit{Women and Life on Earth}. They issued a statement in 1979 calling for a reimagined society founded on ecological principles to prevent the threat of another nuclear disaster. One year later, in March 1980, they organised the first conference on ecofeminism in the US. The event drew over 600 women who engaged in discussions on ecological feminism, peace, and alternative energy solutions. This conference is widely recognised as the moment when the term \textit{ecofeminism} was first introduced in the US, creating a vital intersection between environmentalism and feminism~\cite{Mies1993}.

Forty-eight years later, in 2024, Microsoft announced plans to reopen one of the reactors at the Three Mile Island nuclear power station, the site that originally sparked the ecofeminist movement in the US \cite{Sherman2024}. This intersection of events provides an opportunity to revisit ecofeminist debates in the context of the contemporary race for innovation in Artificial Intelligence (AI). Generative AI (GenAI) and its rapidly growing industry require vast amounts of energy to support the immense computational power needed for model training. GenAI models are built on larger algorithmic architectures, such as neural networks and stable diffusion techniques, which process large datasets to identify patterns and regenerate outputs like images, text, or videos. This computational power is translated into increased energy consumption by the GenAI industry. In fact, the energy demands of GenAI contribute significantly to its industry's carbon emissions~\cite{google2024environmental, Microsoft2024}. These emissions are exacerbated when the power grid supporting the data centres where GenAI models are trained relies on fossil fuels, raising concerns about the tech industry's ability to meet its net-zero commitments. Microsoft, for example, has pledged to eliminate its reliance on diesel fuel by 2030. To achieve this goal, the company has invested in nuclear energy, as exemplified by its plans for the Three Mile Island plant mentioned earlier. Nuclear energy offers a means of reducing dependency on fossil fuels, yet it comes with other environmental consequences such as the generation of radioactive waste, which remains toxic for several generations. 

Beyond carbon emissions, GenAI has additional environmental impacts, including water consumption and waste generation. The GenAI industry relies on vast quantities of potable water to cool the chips used in the training of algorithmic models. For instance, the Large Language Model (LLM) LLaMA-3, developed by Zuckerberg's company, consumed 22 million litres of water over a 97-day period~\cite{Li2023}. To put this into perspective, this is equivalent to the amount of water an average person in England and Wales would use over the course of 424 years.\footnote{According to Statista~\cite{Statista2024a}, one person in England and Wales uses about 142L per day.} E-waste represents another significant environmental issue associated with this technology. It is estimated that the e-waste generated by GenAI could rise substantially in the coming years, amounting to 5.7 million tonnes between 2023 and 2030~\cite{Wang2024}. The environmental impacts of e-waste are primarily linked to its toxic components, such as lead and mercury, which have the potential to contaminate water and soil and pose serious environmental risks to humans and non-humans. As a result, the burgeoning literature on the environmental impacts of AI has emphasised the importance of addressing the sustainability of this technology in the context of the ongoing climate emergency, and thus, of how it could be used as a tool to tackle climate change~\cite{VanWynsberghe2021}. The expanding environmental footprint of the GenAI industry opens a critical opportunity to incorporate ecofeminist perspectives into the debates on its environmental impacts. 

This paper proposes seven principles for \textit{Data Ecofeminism}, aiming to provide critical reflections on how data and AI technologies can be developed nowadays for alternative eco-social transformations. While the environmental footprint of GenAI is contributing to rising carbon emissions and its industry is investing in nuclear energy, these principles offer a theoretical framework that examines the crossroads of intersectional feminism, technology, and environmental issues to promote social and environmental justice. Building on the work of \textit{Data Feminism}~\cite{dignazio2020data_feminism}, the proposed principles draw from ecofeminist, decolonial, and Science and Technology Studies (STS) literature, which have historically traced the connections between science, technology, and discriminatory mechanisms targeting both nature and women. The principles acknowledge that global warming is largely driven by the Global North, with major technology companies exacerbating the crisis by collaborating with the fossil fuel industry. Data Ecofeminism seeks to illuminate the material realities of AI and its supply chains, advocating for transparency by making environmental metrics publicly accessible. It calls for integrating a degrowth perspective---through approaches such as frugal or permacomputing---into AI development, promoting smaller models that perform as effectively as larger ones. The principles also emphasise the importance of digital sovereignty by advocating for autonomous, local, and independent services and infrastructure to meet digital needs sustainably. Data Ecofeminism supports the preservation of physical and digital commons through mutual aid and collective stewardship. And, last but not least, it embraces pluriversal epistemologies, drawing on past and present Indigenous philosophies and their relational understanding of the Earth.

\section{Background}
\subsection{What is ecofeminism?}

Ecofeminism is a movement situated at the intersection of environmentalism and feminism~\cite{herrero2015introductory, Sturgeon1997, bouzo2022amazonas}. It seeks to explore the interactions and potential synergies between the protection of ecosystems and natural resources and the challenges arising from gender inequality. As Vandana Shiva describes it, ecofeminism represents a `new term of ancient wisdom' that `grew out of various social movements---the feminist, peace and the ecology movements'~\cite{Mies1993}. Ecofeminism bridges the gap between environmental, racial, and gender justice by articulating theories and practices that interrogate ideologies that `authorizes oppressions based on race, class, gender, sexuality, physical abilities, and species', while, at the same time also sanctioning `the oppression of nature'~\cite[p. 1]{gaard1993animals_nature}. And in fact, Shiva observed an ideological and material connection between the oppression of women and nature~\cite{agarwal1992gender}. However, this connection has not always been evident. While the term ecofeminism was first introduced in Françoise d'Eaubonne's book \textit{Feminism or Death} (1974)~\cite{deaubonne1974feminism_or_death}, her work was initially met with significant criticism from the French academic community, as they considered women and nature unrelated concepts~\cite{Puleo2013}. Despite this early resistance, her work is now recognised as foundational to the first wave of the ecofeminist movement in the 1970s. 

The first ecofeminist movement, or classic ecofeminism, considered that women are closer to nature based on the ontology of womanhood. This connection arises from the articulation of the feminisation of the environment (\textit{Mother Nature}) and role of giving birth~\cite{brisson2017connectedness}. Given the capacity of women to give birth—though not innate in every woman's body—classic ecofeminists argued that women possessed a unique ability to protect nature. Thus, this first ecofeminist movement advocated for the defence of nature through the maternal connection between nature and the female body. In contrast, the constructivist ecofeminism movement arose in opposition to classic ecofeminism, arguing that the relationship between women and nature is socially and culturally constructed. As scholar and ecofeminist Ynestra King asserted, this connection between women and nature is problematic because it denies women's ability to reason, based on the conventional split between nature and reason~\cite{King1981, brisson2017connectedness}.

Ecofeminism has often been regarded as a predominantly white, Western movement~\cite{Kirk1997, Sandoval1995}, but it has increasingly incorporated decolonial perspectives, drawing on contributions from scholars who examine intersections of feminism, environmental justice, and racial and colonial legacies~\cite{bouzo2022amazonas, gonzalez2020feminismo, Pappucio2018, hajad2024ecofeminism, federicci2022ecofeminism, PerezAguilera2024}. Vandana Shiva critiqued US ecofeminists for overlooking the intersections of Western colonialism, gender oppression, and environmental degradation in the Global South~\cite{VandanaShiva1989}. Bina Agarwal also challenged Western ecofeminism, arguing that it remains a critique without threat to the established order'~\cite[p. 153]{agarwal1992gender} and called for analyses grounded in Global South perspectives~\cite{AzamarAlonso2019}. Agarwal further criticised classical ecofeminism for treating women' as a homogeneous category, failing to account for differences of class, race, and ethnicity~\cite[p. 122]{agarwal1992gender}. Decolonial feminism has raised significant critiques of ecofeminism, emphasising the need to acknowledge the history of colonisation in processes of environmental degradation. For instance, Françoise Vergès, in \textit{A Decolonial Feminism}, calls for a feminism that recognises the legacies of European colonialism and racism, highlighting how racialised women disproportionately bear the burden of managing the waste created by capitalism and extractivism~\cite{Verges2021}. This is evident, as Vergès points out, in examining who cleans train stations in major European cities. In the specific case of AI, this becomes evident when examining who extracts the minerals for electronics, manufactures GPUs, or manages the e-waste generated by AI hardware: usually Indigienous and historically marginalised subjects whos labour and struggles have been invisibilised~\cite{Valdivia2024a, MyersVipra2025, nakamura2014indigenous}.

Despite these differences and critiques, there are three central and common values that are generally embraced in ecofeminism. The first value is plurality, meaning there are as many ecofeminisms as there are theories, viewpoints, and lived experiences~\cite{Puleo2013, gaard1993animals_nature}. In fact, ecofeminism is based in listening to critiques and different perspectives and `tend to pride themselves on the contradictions in their works as a healthy sign of ``diversity'''~\cite{biehl1991rethinking}. Therefore, ecofeminism welcomes plural ideas and diverse perspectives, as evidenced by white and European ecofeminists embracing critiques from decolonial feminists. The second core value highlights the connection between gender inequality and environmental degradation. It argues that the domination of women and the exploitation of nature share a common root cause~\cite{brisson2017connectedness}. Ecofeminists assert that `the subordination of women to men and the exploitation of nature are two sides of the same coin, both responding to a common logic: the logic of domination and the subjugation of life to the logic of accumulation'~\cite[p. 3]{herrero2015introductory}. Finally, the third value is the critique of Enlightenment rationalism and the dominant paradigm of modernity. Ecofeminism thought argues that Western rationalism and modernity have been shaped by masculine values, which have dominated and instrumentalised women's bodies and nature as resources~\cite{Plumwood1993}. These ideologies have imposed values such as biological determinism on sex roles~\cite[p. 3]{Wajcman1991} and deprived nature of any intrinsic value, facilitating its ruthless exploitation and degradation in the name of scientific progress and Western-imposed development~\cite{Merchant1980}. In this vein, Simone de Beauvoir was a pioneer in observing that while women were intrinsically associated with emotions and nature, men were linked to scientific knowledge and reason~\cite{debeauvoir1955pensee, Puleo2013}. This value has been articulated in ecofeminist debates, critically highlighting how Western scientific knowledge has historically designed experiments on women and animals, thereby establishing the inferior status of women and nature~\cite{gruen1993dismantling}. Ecofeminist movements have also pointed out that `technology, like science, is seen as deeply implicated in the masculine project of domination and control of women and nature'~\cite[p. 17]{Wajcman1991}. But how has ecofeminism analysed and conceptualised the role of science and technology?

\subsection{The relationship between ecofeminism, science and technology}

STS have provided crucial scholarly insights into the connection between ecofeminism and technology~\cite{Mies1993, haraway1991simians, King1981, gaard1993animals_nature, Sandoval1995}. As Julia E. Romberger aptly argued in one of the few works addressing the link between ecofeminism and digital technologies: `ecofeminism calls for critical examination of the appropriateness of technology'~\cite[p. 121]{Romberger2011}. That said, ecofeminism should not be seen as an anti-technology or luddite movement. Instead, it should be recognised as a philosophical, theoretical, empirical, and political praxis that: (1) recuperates science and technology from feminist, decolonial, and environmental perspectives~\cite{Sandoval1995}; (2) challenges any logic of domination, capitalist modes of production, and the hegemony of mechanistic science~\cite{haraway1991simians}; and (3) considers heterogeneous perspectives from those historically marginalised, with the aim to `build a world of many worlds where everybody fits'~\cite{delacadena2018worlds, Sandoval2000}.

Ecofeminism has historically criticised technocratic approaches~\cite{birkeland1993ecofeminism} and technologies that threaten life on Earth~\cite{Mies1993, ShivaHerrero2024}. Since Carson's \textit{Silent Spring} (1962)~\cite{carson1962silent} demonstrated the environmental harms caused by pesticides used by soldiers during WWII, ecofeminists have challenged the illusion of rational and objective technology. They have shown how technological artifacts are shaped by Western, capitalist, and masculine values, which are used to dominate both nature and women's bodies~\cite{cockburn1981male_power}. In this vein, Haraway also argued in \textit{Simians, Cyborgs, and Women} (1991) that after WWII, `electronic industries and communications technology were increasingly tied to strategies of social and military planning'~\cite[p. 58]{haraway1991simians}. These ecofeminist criticisms have also focused on production, reproductive, and domestic technologies~\cite{Wajcman1991} as well as militarist and nuclear technologies~\cite{birkeland1993ecofeminism}. More recently, ecofeminists in Latin America have questioned sustainable development technologies, such as green energy projects, arguing that they also contribute to environmental harms and serve as a rationale for extending capitalist activities~\cite{Isla2021, Phillips2020}. However, little attention has been given to the connection between ecofeminism and data and AI technologies. Given the increasing environmental footprint of GenAI technologies and their associated industries, this paper proposes an ecofeminist approach to data and AI technologies such as ChatGPT and contributes to the FAccT critical literature by opening a dialogue between ecofeminism and GenAI technologies.

\section{Related work}

\subsection{On the environmental impacts of data and AI technologies}

In recent decades, AI has been promoted as a key tool for tackling climate change~\cite{cowls2023ai_gambit, coeckelbergh2021ai_climate}, with claims such as `AI is essential for solving the climate crisis'~\cite{Maher2022}. Yet, critical analyses of the environmental costs have raised doubts about the sustainability of AI~\cite{dobbe2019ai_climate, brevini2021ai, VanWynsberghe2021}.  Recent developments in Deep Learning (DL) are considered computationally intensive, meaning they require vast amounts of energy and water to train algorithmic systems.\footnote{The electronic circuits used to train GenAI, such as GPUs, NPUs, and TPUs, require more electricity than previous circuits to analyse data and train algorithms. Increased electricity consumption also translates into higher water usage to cool these chips.} If this energy comes from burning fossil fuels, training algorithmic systems results in carbon emissions, which is widely recognised as an environmental impact~\cite{hogan2024fumes, Luccioni2023}. Within DL, GenAI models have raised significant concerns about the benefits of AI technologies in addressing climate change, due to their computational costs and the resulting environmental impacts. Indeed, Google, Meta, and Microsoft have acknowledged in their sustainability reports that GenAI technologies are driving up their environmental impacts. As a result, the growing literature on the environmental effects of GenAI has shown that, rather than being a solution to climate change, this technology is contributing to it~\cite{crawford2024generative_ai, Stein2024, brevini2021ai, Lehuedé2024, hogan2015data, brodie2023data, Kneese2024}.

Within the FAccT community, Emily Bender et al. (2021) were pioneers in raising awareness about the environmental harms of LLMs in their influential work, on \textit{Stochastic Parrots}~\cite{bender2021stochastic}. As a result, FAccT scholarship has increasingly focused not only on the social and technical dimensions of AI but also on its environmental impact~\cite{Rakova2023}. As a result, taxonomies of algorithmic harms are now incorporating ecological harms as part of the risks associated with AI implementation, such as carbon emissions and water usage~\cite{Kneese2024, dodge2022carbon_ai}.  In this context, Luccioni, Viguier, and Ligozat (2023) investigated the carbon footprint of large models such as BLOOM, unveiling that it emits 50 tonnes CO$_2$eq emissions~\cite{Luccioni2023}. As this body of work demonstrates, examining the infrastructure of GenAI technology reveals harms that data centres pose not only to the environment but also to local communities. For example, media scholar Mél Hogan (2015) highlighted that data centres in the US extract large amounts of water~\cite{hogan2015data}, while Libertson et al. (2021) and Bresnihan and Brodie (2021) criticised the substantial energy demands of the data centre industry, which disrupts energy access and has led to electricity grid collapses in local communities in Sweden and Ireland, respectively~\cite{Libertson2021, bresnihan2021extractive}. Furthermore, other critical scholars have argued that the environmental harms of AI should be considered across the entire production process, rather than focusing solely on data centres. Examining other industrial activities crucial to AI development, such as mineral extraction, chip manufacturing, and e-waste dumping, uncovers additional environmental harms~\cite{Valdivia2024a, Taffel2021}. These perspectives suggest that attention to the environmental harms of AI should also encompass issues such as soil degradation and the waste generated by the AI industry, extending beyond carbon emissions and water withdrawal just during the algorithmic training phase.

\subsection{An ecofeminist approach towards data and AI}

Feminism and AI are two interconnected concepts that have been widely analysed, but what about ecofeminism and AI? There have been numerous studies highlighting the need to incorporate feminist perspectives into AI. As an example, the \textit{Data Feminism} framework proposed by scholars D'Ignazio and Klein (2020)~\cite{dignazio2020data_feminism} draws on seven principles to expose how feminism can help rebalance power in AI/ML research, including embracing pluralism, which is strongly aligned with ecofeminist values. However, three years after their book was published, they revisited these principles, emphasising the importance of analysing the environmental issues posed by AI development and deployment, which should be integrated into data feminist approaches~\cite{Klein2024}. Despite this, AI feminist perspectives have not yet explored the connection between gender oppression and environmental injustice in the technological context. 

As discussed earlier, an ecofeminist perspective on AI must encompass its values, such as plurality, acknowledge the connection between gender and environmental struggles, and critique the hegemony of Western science. This latter ecofeminist value has been crucial in critical feminist and STS scholarship, such as the work of Judy Wajcman, Ruha Benjamin, and Donna Haraway, who have shed light on how technological artifacts embody white and masculine hegemonic values and ideologies~\cite{Wajcman1991, benjamin2019race, haraway1991simians}. For instance, in the early 1990s, Haraway discussed how `Operations Research began with the WWII and efforts to coordinate radar devices… Statistical models were increasingly applied to problems of simulation and prediction for making key decisions'~\cite[p. 58]{haraway1991simians}. This body of scholarship led critical scholars to recognise that AI technologies, which are based in Operational Research and Statistics, among other fields, also reproduce white and masculine epistemologies~\cite{gebru2024tescreal, benjamin2019race, crawford2016white_guy, birhane2022values, bolukbasi2016debiasing} and show connections to the US military-industrial complex~\cite{Katz2020, TheNewYorkTimes2021}. In fact, OpenAI has recently announced contracts to deploy AI in the battlefield~\cite{MITTechReview2024}. During the Radical Book Fair held in Barcelona, Yayo Herrero and Vandana Shiva (2024) discussed the past and future of the ecofeminist movement. Yayo Herrero, an Spanish scholar and ecofeminist, criticised the use of AI in military operations, emphasising that through the AI industry, we are witnessing a `new wrinkle in a technology of death and war, which is able to detect whose life is dispensable'~\cite{ShivaHerrero2024}. Herrero referred to \textit{Lavender}, a Machine Learning (ML)-based algorithmic system implemented by the Israeli army to identify and kill individuals allegedly linked to Hamas~\cite{abraham2024lavender}. With a 10\% error rate, this system was deployed to target missiles at individuals previously detected by the algorithm, which could result in the deaths of innocent people. This algorithm directly challenges ecofeminist epistemologies, which prioritise the value of all life, human or non-human, on Earth. Moreover, the increasing environmental harms caused by GenAI-intensive computing present an opportunity to integrate an environmental perspective into feminism and AI. This paper responds to that call by drawing on \textit{Data Feminism} and proposing seven principles that add to them and guide us toward \textit{Data Ecofeminism}.

\section{Data Ecofeminism}

The seven principles for \textit{Data Ecofeminism} emerge from the need to establish data and AI research guidelines---particularly for GenAI technologies---that align with intersectional and decolonial feminism, while remaining critically aware of the environmental degradation and commodification of nature that making AI entails. This technology, which continues to reproduce existing algorithmic inequalities such as gender, race, and class bias, is also contributing to climate change by exacerbating environmental impacts. These principles are proposed to guide the development of digital data technologies and algorithmic systems that challenge technological power, redirect it, and help build a thriving society within a living, caring world where everyone---human or non-human---can fit.

\subsection{Principle 1: Examine the climate emergency and its power structures}

\textit{Data Ecofeminism recognises that wealthy countries are significant contributors to global warming.}

According to the IPCC (2022)~\cite{ipcc2022climate}, it is unquestionable that human and capitalist activities have warmed the atmosphere, contributing to global warming and increasingly extreme weather events and natural disasters. Capitalist societies have accelerated the climate crisis by accumulating capital through the extraction of resources that commodify nature. However, asymmetric and unequal power relations emerge in this context, giving rise to environmental injustice. High-income countries bear a substantially higher degree of responsibility for climate change, while low-income countries suffer disproportionately more from its natural disasters, despite not contributing to it~\cite{Oxfam2023}. While the Global North was responsible for 92\% of excess global CO$_2$ emissions~\cite{hickel2020quantifying}, more than 91\% of deaths caused by climate change occurred in the Global Majority~\cite{UN2021}. The US and EU-28 countries are at least twice as responsible as China or India, given their historical contributions to carbon emissions. But it is not only about carbon emissions; a recent publication in Nature Sustainability illustrated that minerals and metals essential for manufacturing electronics are primarily extracted from Indigenous and peasant lands~\cite{Owen2023}. In fact, the political and power structures in the context of climate change are entangled in a colonial and racial legacy that imposed its capitalist rationality around the world~\cite{fraser2023cannibal, hickel2022less}.

Given these disparities, ecofeminist movements have called for the politicisation of climate change by understanding how power structures within capitalism are orchestrated~\cite{herrero2023sustainability}. In the words of the decolonial feminist and scholar Françoise Vergès: `Global warming and its consequences for the peoples of the South is a political question and must be understood outside the limits of ``climate change'' and in the context of racial capital'~\cite[p. 7]{Verges2017}. Therefore, being aware of climate change and environmental injustices also encompasses comprehending different vectors of struggle, such as class, gender, race, and nationality, among others.
 Within this context, the AI industry has been governed by white and male elites whose Western, colonial and capitalist ideologies have also been embedded within algorithmic models~\cite{crawford2016white_guy, Katz2020}. Digital data is a crucial factor in the development of AI and GenAI, and numerous studies have demonstrated how the big tech industry reproduces these ideologies—for instance, see Couldry and Mejías (2019) for \textit{Data Colonialism}~\cite{couldry2019costs} or D'Ignazio and Klein (2020) for \textit{Data Feminism}~\cite{dignazio2020data_feminism}. From an ecofeminist perspective, the AI industry is also contributing to climate change by courting major fossil fuel companies and offering AI-based products that facilitate extraction~\cite{dobbe2019ai_climate}. According to Greenpeace's report \textit{Oil in the Cloud} (2020), Microsoft, Google(Alphabet), and Amazon have contracts with the oil and gas industry to unlock oil and gas deposits using ML~\cite{greenpeace2020oil}. Another AI company, in this case, NVIDIA, is also partnering with Petrobras to explore new oil and gas extraction fields through AI~\cite{NVIDIA2024}. Moreover, the carbon emissions of private jets used by big tech billionaire CEOs are also contributing to climate change: `The two private jets owned by Jeff Bezos, founder and executive chairman of Amazon, collectively spent almost 25 days in the air, emitting 2,908 tonnes of CO$_2$. It would take the average US Amazon employee almost 207 years to emit that much'~\cite[p. 9]{Oxfam2024}. 
 Thus, a data ecofeminist approach calls for a critical examination of the economic and corporate male elites who instrumentalise their power in the AI industry to accumulate capital through natural resource exploitation. This approach considers that the AI industry, which is predominantly based in the Global North, is also contributing to climate change, in some cases even more than entire nations based in the Global Majority~\cite{forbes2023climate}.

\subsection{Principle 2: Consider digital materiality and its supply chains}

\textit{Data Ecofeminism acknowledges that datasets and GenAI infrastructure are derived from natural resources, emphasising their supply chains and associated environmental impacts.}

Narratives about the so-called `cloud' have obfuscated the materiality of massive digitalisation. As media ecology scholar Sy Taffel claimed, attention should be drawn to the material aspects of the `planetary-scale extractive industries' developed for the acquisition of digital data~\cite{Taffel2021}. The cloud, or data farm, is a concrete infrastructure, also known as a data centre, that contains many servers stored in rooms that process data and allow our Internet connections~\cite{hristova2022datafarms}. Over the last decade, the number of data centres has increased to power the digitalisation of our daily lives. This industry is becoming one of the most profitable in the digital economy~\cite{cancela2023utopias, datacentre2024investment, brodie2023data, Lehdonvirta2022}, but its economic activity has a massive environmental footprint~\cite{Whitehead2015}. Yet, these narratives about the cloud do not only obfuscate its materiality~\cite{Mills2024}, but also its environmental impact~\cite{devries2023ai_energy}. Most data centres use fossil fuels, such as gas, oil, or coal, to power their servers, which contributes to a surge in carbon emissions by big tech companies~\cite{google2024environmental, Microsoft2024}. Data centres also require vast amounts of water to cool down server rooms, causing struggles in local communities~\cite{Lehuedé2024, brodie2023data}. In some cases, the data centre industry uses vast amounts of water in regions suffering severe droughts, where local communities do not even have access to drinking water in their own houses~\cite{Valdivia2024a}. Therefore, the materiality of GenAI and its environmental impact have been recently scrutinised by scholars, journalists, and environmental activists, given the environmental and social challenges they entail.

\begin{figure}[h]
  \centering
  \includegraphics[width=0.8\linewidth]{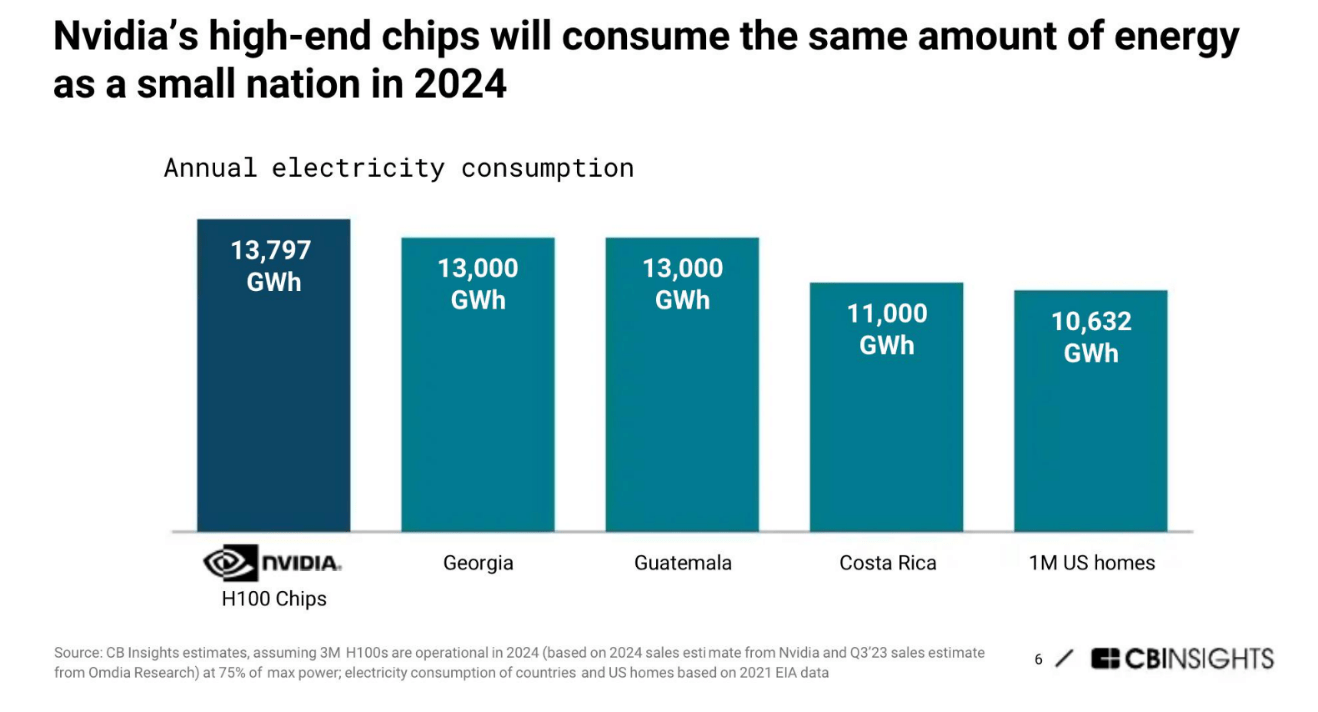}
  \caption{NVIDIA's chips for GenAI consume more energy than Georgia or Guatemala~\cite{cbinsights2023nvidia}.}
  \label{fig:nvidia}
\end{figure}

The increasing size of GenAI models has transformed the AI industry: it is estimated within 5 years the size of NLP models has increased by 15,000 times~\cite{AWS2024}. Given the size of these models, GenAI technologies need to be trained in data centres with GPUs, which is accelerating the construction of more data centres that are in turn demanding for more chips~\cite{Reuters2024}. However, examining GenAI materiality also encompasses drawing attention to its entire supply chains which are `the orchestration of commodity chains that extract, ship, and manufacture the natural resources needed to develop AI from an infrastructural perspective, such as mines, data centres and e-waste dumps together with their human resources'~\citep[p. 5]{Valdivia2024a}. Within these supply chains, the tech company that provides GPUs worldwide, NVIDIA, has ranked first in the list of market capitalisation, surpassing other big tech companies such as Apple, Facebook(Meta) or Google(Alphabet). With an increase in its market value, this suggests that NVIDIA could supply more chips for GenAI, which would entail the extraction of additional natural resources. It is estimated that the energy used by NVIDIA's chips in 2024 is greater than the energy used by Georgia or Guatemala (see Figure~\ref{fig:nvidia}). Within this supply chain, the Taiwanese firm, TSMC, which is the semiconductor firm that manufactures and provides chips to NVIDIA, is becoming a key actor within the AI industry providing almost the 80\% of chips used for AI training. Given the importance that this semiconductor factory has gained in the global economy, the Taiwanese government notified to rice farmers that water should be prioritised to TSMC during a drought episode~\cite{TheNewYorkTimes2021}. A data ecofeminist approach considers the material dependency of digital technologies and the ecological harms and social impacts that emerge across these chains.

\subsection{Principle 3: Report environmental metrics}

\textit{Data Ecofeminism proposes to make publicly available environmental metrics and effectively reduce them.}

In 2020, The Register published a report claiming that ChatGPT's foundational model, GPT-3, had the same carbon footprint as a 700,000 km car trip, which is twice the distance between the Earth and the Moon. An interesting aspect of this report was the quantification of algorithmic carbon emissions. For instance, while a classic NLP pipeline is estimated to emit 78,468 lbs of CO$_2$, a transformer model is estimated to emit 626,155 lbs~\cite{Strubell2020}. GenAI systems require more computing power than classic ML algorithms because they can handle more data and perform more complex calculations compared to their algorithmic predecessors. More recently, Luccioni, Viguier, and Ligozat (2023) also estimated the carbon footprint of a 176-billion parameter language model, which emitted approximately 50 tonnes of CO$_2$ taking into account manufacturing and training activities~\cite{Luccioni2023}. However, these estimations only consider carbon emissions during the algorithmic training phase. A data ecofeminist perspective should ask: What are the carbon emissions of GenAI technologies when we consider the entire supply chain?

\begin{figure}[h]
  \centering
  \includegraphics[width=\linewidth]{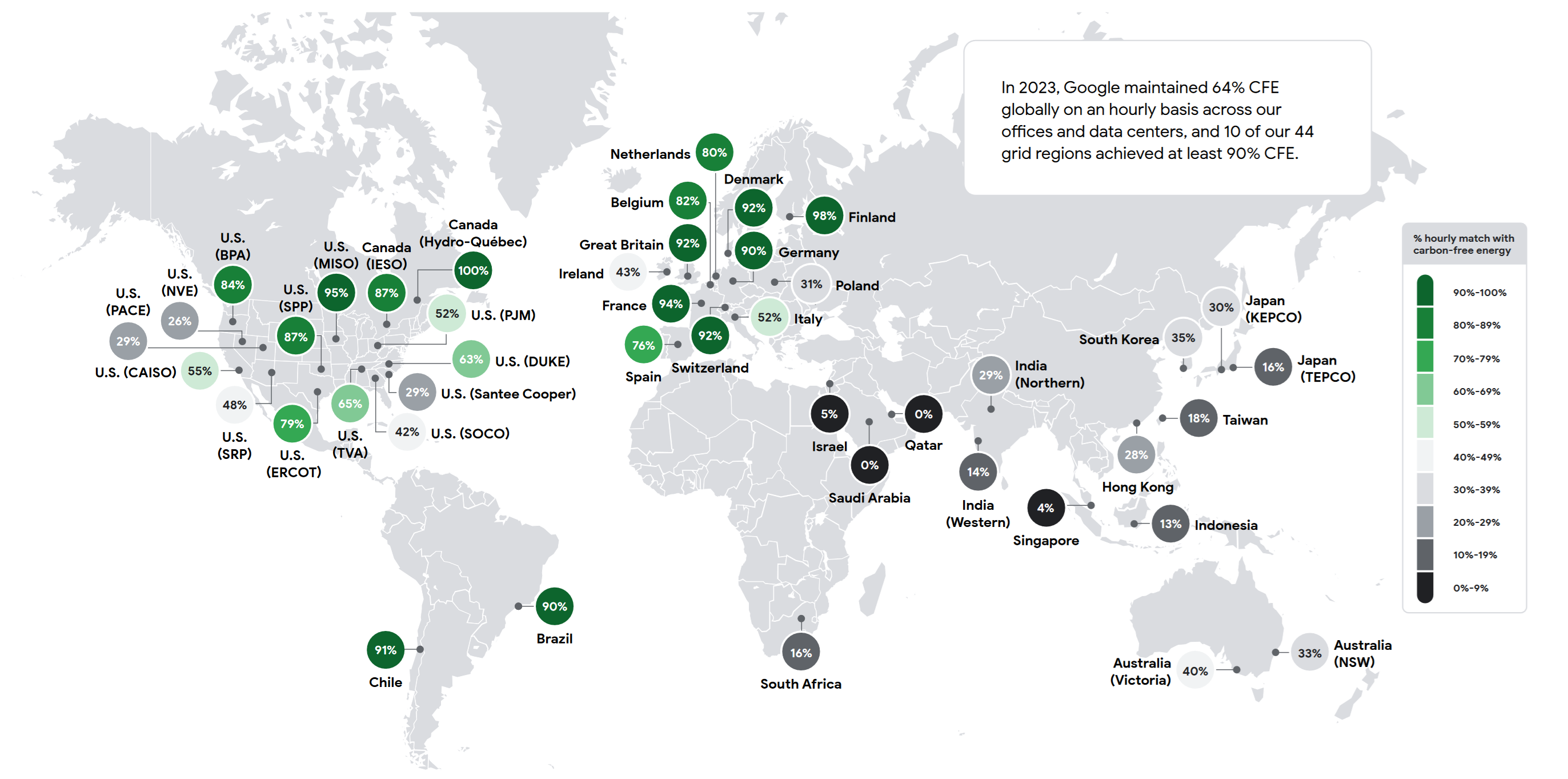}
  \caption{Google's data centres are generally more sustainable in the Global North~\cite{google2024environmental}.}
  \label{fig:google}
\end{figure}

While Principle 2 proposes considering the materiality and supply chains of GenAI technologies, Principle 3 advocates for making the environmental metrics associated with this process publicly available. Most big tech companies report their environmental metrics annually through Corporate Sustainability Reports (CSR). For instance, Google's CSR illustrates that data centres are generally more sustainable in the Global North, meaning that Global Majority nations are more exposed to the environmental costs of GenAI infrastructure (see Figure~\ref{fig:google}).

Yet, these sources do not offer a detailed estimation of the carbon footprint of their suppliers. Extracting minerals, manufacturing chips, building and making operational a data centre, executing algorithms, generating and gathering data, storing datasets, and finally dumping electronics—all of these activities entail carbon emissions and should be taken into account when estimating algorithmic carbon emissions. However, it is a challenge to account for carbon emissions within supply chains because firms do not hold their providers and suppliers accountable~\cite{Stenzel2023}. For instance, while NVIDIA publishes their carbon emissions annually in the CSRs, there is a lack of detailed information on the carbon emissions from mineral extraction to manufacture GPUs.

In this context, the European Parliament and the European Commission have introduced two pieces of legislation to make environmental information more accessible. On one hand, the Energy Efficiency Directive 2023/1791 (2023) stipulates that `[t]he Commission shall establish a European database on data centres that includes information communicated by the obligated data centres. The European database shall be publicly available on an aggregated level'~\cite{TheEuropeanParliament2023}. This regulation was enforced in May 2024, but was delayed to September 2024. However, following a Freedom of Information petition designed and sent by the author of this piece, the European Commission replied that this database is not publicly available yet. On the other hand, the Corporate Sustainability Directive 2024/1760 (2024), also known as the Supply Chain Act, establishes core elements to address adverse human rights and environmental impacts in a company's own operations, its subsidiaries, and, where related to their value chains, those of their business partners~\cite{TheEuropeanParliament2024}. Among other obligations, this legislation requires companies to reduce greenhouse gas emissions across their entire supply chain to combat climate change.

Apart from carbon emissions, there are other environmental metrics that should also be made publicly available within the GenAI industry, such as water withdrawal and discharge or waste generation. The environmental costs of GenAI technologies are not limited to carbon emissions, but also include intensive water usage and the amount of e-waste generated. For instance, as scholars Kneese and Young (2024) have claimed, `one problem with only considering the carbon emissions associated with LLMs and ignoring other environmental impacts is that optimizing for reducing the carbon emissions of training a model may actually exacerbate the water cost'~\cite{Kneese2024}. A data ecofeminist approach calls for making all these metrics publicly available so that companies can be held accountable and systematically reduce their environmental metrics, thereby helping to limit global warming to 1.5$^{\circ}$C in line with the Paris Agreement~\cite{UN2015} and respecting planetary boundaries, as well as labour, environmental, and other fundamental rights.

\subsection{Principle 4: Prioritise frugal AI computing}

\textit{Data Ecofeminism adopts a degrowth perspective towards data and AI technologies by using less energy, resources and materials.}

Data and GenAI is accelerating growth driving up big tech's environmental footprint~\cite{Meyers2023}. According to Statista, in 2024 the world generated 149 zettabytes of data, and it is projected to generate 394 zettabytes by 2028 (2024b)~\cite{Statista2024b}. The generation and consumption of data contributes to increase energy demand: the more data that is generated, the more energy is used. At the same time, this data is also used to feed GenAI models that also contribute to this energy demand. As a result, Microsoft increased its carbon emissions by 40\% during the 2020-2023 period due to ChatGPT and Copilot, Facebook (Meta) saw a 65\% increase in emissions from 2021 to 2023, and Google (Alphabet) reported a 50\% higher carbon footprint in 2023 compared to 2019, attributing this `due to increasing energy demands from the greater intensity of AI compute'~\cite{TheConversation2024}. GenAI is also poised to drive 160\% increase in data centre power demand~\cite{goldman2024ai_power}. As an example, Amazon has announced that their three data centres in Aragón (Spain) are going to use more energy than the whole region~\cite{LaMarea2025}. And water, it is estimated that the AI industry is going to take between 4.2 and 6.6 billion cubic meters of fresh water~\cite{Li2023}, approximately 4- and 6-times Denmark's water withdrawal. GenAI is also requiring more elements~\cite{Parikka2021}: GPUs are nowadays using more chemical elements than chip predecessors in the 80s~\cite{Mitra2024}. And waste, GenAI is also generating more e-waste and could increase 1000 times electronic residues, from 2,600 tonnes in 2023 to 0.4-2.5 million tonnes in 2030~\cite{Wang2024}. This growth in energy, water withdrawal, carbon emissions, materials extracted, and waste generated is clearly impacting on climate change and this data ecofeminist approach proposes to embrace a degrowth perspective towards computing to mitigate its environmental impact.

Degrowth is defined by scholar Jason Hickel as a `planned reduction of excess energy and resource use to bring the economy back into balance with the living world in a safe, just and equitable way'~\citep[p. 29]{hickel2022less}. Hickel together with other degrowth scholars such as Alice Mah, Serge Latouche, Carlos Taibo and Giorgios Kallis based in the Global North propose that degrowth is the only alternative to mitigate climate change, empower environmental justice and create a sustainable future. But how is a degrowth perspective translated into the data and GenAI context? In computing, digital degrowth is recognised as a form of sustainable computing~\cite{Selwyn2023} and has introduced the concept of frugal computing, which aims to `reduce emissions from computing by using less energy and less materials'~\citep[p. 4]{vanderbauwhede2023frugal}. Within this scenario, there is also the interesting perspective of permacomputing~\cite{heikkila2020permacomputing, Sursiendo2022} that `asks the question whether we can rethink computing in the same way as permaculture rethinks agriculture'~\cite{Miyazaki2022}. A degrowth and frugal perspective towards computing challenges the ever-expanding resource and energy use of data and GenAI technologies while questioning the desirability of algorithmic products. 

Concretely, from a software perspective, it implies using small models than performs similarly to larger models~\cite{chen2023frugalgpt}, building models with a smaller number of features without compromising performance~\cite{Narayanan2019} and storing less data. From a hardware perspective, it entails producing less chips, building less data centres, and generating less e-waste, among other measures~\cite{cancela2023utopias}. An ecofeminist approach towards AI challenges the construction of sustainable data centres due to the Jevon's Paradox: the construction of more energy efficient data centres increases rather than reduce energy consumption. Another crucial aspect within the degrowth scenario is the \textit{Right to Repair}, a global movement that aims at extending lifespan of products to reduce waste. In the context of AI, while it is estimated that a GPU has an average lifespan of 5 years~\cite{Valdivia2024a} which could dramatically contribute to e-waste global issue, a degrowth perspective calls for extending this lifespan and allow GPU users to repair their electronics.

\subsection{Principle 5: Reclaim digital sovereignty}

\textit{Data Ecofeminism calls for a public-led digital infrastructure.}

In 1991, Donna Haraway observed that `the new technologies seem deeply involved in the forms of ``privatization'' [...] in which militarization, right-wing family ideologies and policies, and intensified definitions of corporate (and state) property as private synergistically interact'~\citep[p. 168]{haraway1991simians}. Thirty-four years later, in 2024, this same observation applies to big tech companies, given their political and economic dominance in the AI sector and the companies that control it. Google(Alphabet), Facebook(Meta), Microsoft, and Amazon account for 65\% of global Internet traffic~\cite{Sandvine2024}, much of which traverses their owned subsea cables---now comprising 71\% of the total, a substantial rise from just 10\% a decade ago~\cite{IndoPacific2024}. These companies are also estimated to control a significant share of global datasets, which are essential for training their AI-based models. They develop and maintain digital platforms, such as email services and data storage, which both governmental and non-governmental actors heavily rely on. For instance, UK universities use Microsoft cloud and email services, while UK government departments and public bodies depend on Google's email services and Amazon's cloud storage. This dominance has positioned big tech firms at the forefront of the AI innovation race, with the latest GenAI models being developed primarily by the private sector rather than public universities.

Given the infrastructural, economic, and political influence big tech companies are accumulating, there have been calls for digital sovereignty. This concept refers to a government's ability to act independently in the digital `should be able to take autonomous actions and decisions regarding their digital infrastructures and technology deployment'~\citep[p. 8]{Pohle2020}. A recent policy paper on digital sovereignty proposed a `progressive reform agenda to enhance digital sovereignty for people and the Planet'~\citep[p. 2]{rikap2024reclaiming}. This agenda is built on four measures: offering a democratic and public digital infrastructure, designing a research agenda that is not driven by technological hype and solutionist narratives, promoting a public knowledge network that challenges profit- and control-motivated models, and expanding human and civil rights. A key aspect of this proposal, aligned with an ecofeminist approach, is its emphasis on prioritising life over market profit. In other words, it places people, the Planet, and democracy above private economic gains. For example, during severe droughts, the data ecofeminist approach  prioritise water resources for essential activities that ensure life for human and non-humans.

Building on digital sovereignty frameworks, a data ecofeminist approach envisions the creation of a public-led digital infrastructure that addresses people's essential digital needs while minimizing environmental impacts. This vision emphasizes the need to build a public cloud that eliminates dependency on private, profit-driven providers, thereby shifting the focus from economic gain to public welfare. It also advocates for the development of technological solutions that prioritize the well-being of people and the planet. A key aspect of this approach is dismantling practices of illegal data extraction and labor exploitation within the digital ecosystem. For instance, concerns have been raised about LLMs being trained on copyrighted materials, infringing upon the rights of writers and artists, and potentially undermining their livelihoods~\cite{ArteEsEtica2023}. Furthermore, these models often rely on low-cost and exploitative human labor, presenting significant challenges to labor rights~\cite{Muldoon2023}. To align with an ecofeminist framework, digital sovereignty must adhere to legal and ethical standards, actively working to avoid perpetuating or reconfiguring existing social and labor injustices. This approach also necessitates a critical evaluation of which AI products are truly essential to develop, particularly in the context of the climate crisis. Ensuring that technological innovation aligns with ecofeminist values is vital for building a sustainable and equitable digital future.

\subsection{Principle 6: Prioritise the commons and mutual aid}

\textit{Data Ecofeminism reclaims the communality in the digital era.}

Communality refers to a shared sense of mutual aid~\cite{Kropotkin1902}, rooted in common interests and goals. It is supported by practices and services grounded in the values of reciprocity, participatory democracy, sustainability, and activities that prioritize life, people, and the planet~\cite{barcena2023prologue}. Communality thrives through the commons—cultural and natural resources that are collectively accessible and held in trust for mutual aid and social reproductive activities. Proposed as a strategy to mitigate the consequences of the climate crisis~\cite{barcena2023prologue}, communality ensures access to food and other basic needs~\cite{federicci2022ecofeminism} in a sustainable way~\cite{euler2019commons} while fostering mutual aid during crises. For example, during the flash floods in Valencia, Spain, in October 2024—caused by unusually warm Mediterranean waters linked to global warming—mutual aid among citizens played a crucial role in addressing the immediate consequences. Despite the tragedy of 224 fatalities, community efforts were instrumental in mitigating the disaster's impact in its early days~\cite{LaParadoja2024}. Historically, the commons were vital to survival, particularly in Europe during the late Middle Ages, where pastures, forests, and rivers were shared by peasants to grow crops, access water, and sustain communities~\citep[p. 46]{hickel2022less}. Scholars have noted that the earliest phase of capitalist development was marked by the dispossession and privatization of these shared resources~\cite{Angus2023, Kropotkin1902, Mies1993}. But how can the concept of communality be reimagined in the digital sphere through an ecofeminist lens?

Ecofeminism and the commons are deeply interconnected, as women have historically played a central role in defending their territories and asserting their `right to land'~\cite{federicci2022ecofeminism}. Similarly, the Internet—originally conceived as a public resource—has been increasingly appropriated by the predominantly male-dominated big tech sector. In response, digital feminist movements have risen to reclaim and reimagine digital spaces. Decades ago, cyberfeminism emerged as a counterpoint to male-dominated visions of technology~\cite{Plant2013, Schaffer1999}. Today, new forms of digital feminist communities continue to empower historically marginalised groups, emphasising care and community as core values~\cite{franco2024networks}.

The concept of digital commons—defined as `digital resources that are commonly controlled by humans'~\citep[p. 19]{fuchs2021digital_commons}---offers a promising alternative for providing basic public services and fostering participatory democracy. Examples include public subsea cables, non-profit data centers, and free and open-source software~\cite{cancela2023utopias}, which can also be more sustainable than private ownership models~\cite{euler2019commons}. As GenAI technologies and their associated industries drive increasing consumption of energy, land, data, and water—resources often exploited for private interests—a data ecofeminist approach advocates for the creation and reappropriation of both physical and digital commons. This perspective calls for a critical examination of GenAI's impact on communal resources~\cite{creativecommons2023ai_commons}, the protection of communal land against expropriation for profit-driven activities, and the development of ecofeminist infrastructures that support life, people, and the planet~\cite{hache2023feminist_infra}. A data ecofeminist framework envisions nurturing relational worlds of care both within and beyond the digital sphere. It advocates for the development of data and AI technologies that embrace mutual aid and prioritize tools essential for advancing social and environmental justice, particularly in the face of the climate crisis. By reclaiming communality in the digital era, data ecofeminism seeks to build technologies that serve people and the planet rather than perpetuate exploitative systems of power and capitalist profit.

\subsection{Principle 7: Weave the pluriverse}

\textit{Data Ecofeminism values historical and current Native epistemologies to build a world where many worlds fit.}

In \textit{Azmapu} (2023), a piece written by Elisa Loncon, the author revisits the values of Mapuche philosophy related to the care of the Earth~\cite{Loncon2023}. One of the key points within this philosophy, which is also common in many other Indigenous philosophies, is the understanding that the world is interrelated, meaning that every element within it is connected. As such, Loncon explains that Mapuche philosophy does not consider natural resources as such, which is a Western and anthropocentric concept. Instead, this philosophy acknowledges all beings of nature and develops a reciprocal relationship to foster care and respect. By doing so, the Mapuches cultivate a connection with nature, fostering a deep commitment to protecting their lands and preserving their biodiversity from capitalist extractive projects~\cite{Nesti2001}. This is also what the decolonial scholar Arturo Escobar has introduced as radical relationality: `the fact that all entities that make up the world are so deeply interrelated that they have no intrinsic, separate existence by themselves'~\citep[p. xiii]{escobar2020pluriversal}. Escobar proposes that radical relationality emerges as a counterproject to the modern and capitalist struggles for social and environmental justice. Yet, radical relationality and Native epistemologies are often dismissed by the hegemonic project of modernity, which is rooted in Enlightenment rationalism and Western science, and, as ecofeminists have criticised, it has historically been dominated by male-dominated cultural institutions that have devastated the planet~\cite{gruen1993dismantling}.

The data ecofeminist approach aims at weaving the \textit{pluriverse}, or in other words, to embrace `the heterogeneous worldlings coming together as a political ecology of practices'~\cite[p. 4]{delacadena2018worlds}, by acknowledging the radical relationality among human and non-human beings and the right of rivers, forests, and seas to be safe from capitalist and polluting activities. This aim builds on the fact that ecofeminism is considered a plural movement, where there are as many ecofeminisms as theories, viewpoints, and vital experiences, including Native cosmologies~\cite{Puleo2013, gaard1993animals_nature}. Within the pluriverse, the data ecofeminist perspective proposes that Native epistemologies should have a relevant role as a reparation action given the historical discrimination that Native peoples have suffered in their efforts to protect their land and epistemologies against colonialism, evangelization, and Western-imposed development~\cite{escobar2018pluriverse}. Moreover, \textit{Data Ecofeminism} reconsiders AI impacts on the environment through radical relationality among human and non-human beings, seeing beyond natural resources, which embraces other historical and  contemporary Native epistemologies, such as that of the Mapuches~\cite{escobar2018pluriverse}. In other words, an ecofeminist relationship between AI and the environment should not only envision how GenAI technologies are damaging the environment through quantitative estimations of carbon emissions and water withdrawal measurements but also acknowledge different perspectives and experiences of how natural resources are seen as other beings, based on interconnectedness, and hold their own rights to be protected and preserved.

\section{Discussion}

Are there ways of developing data and AI technologies from an ecofeminist perspective?
Data ecofeminism is a theoretical and analytical framework made up of seven principles that aim to respond to this question by delineating the pathway toward developing data and AI technologies that put \textit{care} at their core, while respecting life, people, and the planet. While the accident at the Three Mile Island nuclear power station in 1979 set the precedent for the first ecofeminist group and conference in the US, the reopening announcement of this nuclear station in 2024 to power GenAI technologies invites a reconsideration of how ecofeminist epistemologies confront GenAI technologies and their industry---similarly to how feminism confronted technologies decades ago~\cite{Wajcman1991}. With the growing resource-intensive technologies like GenAI, which are putting into question how AI could tackle climate change, the \textit{Data Ecofeminism} principles offer seven dimensions of theoretical and practical guidance to critically advocate for sustainable technologies. These principles, rooted in the world-leading work of Data Feminism (2020), integrate an ecofeminist perspective toward AI and data science by bridging the gap between social and environmental justice. Rather than merely accounting for carbon emissions and water usage of AI, data ecofeminism calls for remaining critically and politically aware of the structural and historical mechanisms that sustain this technology today, offering a guide to develop radical alternatives that keep the Earth's temperature rise to 1.5$^{\circ}$C, in line with the Paris Agreement~\cite{UN2015}.

Ecofeminist movements and critical feminist STS scholars have commonly pointed out that Western science and technology, rooted in the Enlightenment and the logic of reason, has been imposed to dominate the body and the territory, women, and nature. In doing so, it has also disrupted and commodified nature for market profit~\cite{Mies1993}. As an example, military technologies such as nuclear power and weapons have been at the forefront of ecofeminist critiques, illustrating how these technologies are tools of `death' rather than `life' given the negative consequences that nuclear energy has on the environment and human health. Given the current trend of AI development, with the big tech industry investing in nuclear reactors to power GenAI systems alongside the large number of resources needed to train these models, this paper proposes bringing ecofeminism into the critical AI scholarship, not only to illustrate how feminist values could be useful in building technology but also to guide us in building technology that `does not reproduce an exploitative relationship with nature and ourselves'~\citep[p. 59]{Wajcman1991}.

Data ecofeminism confronts data and AI epistemologies, and more recently, GenAI technologies, which are presented as the latest technocentric experiment devastating nature~\cite{Ricaurte2019}. However, this ecofeminist approach is not against technology, because `ecofeminism believes that critical technology use involves careful, context-specific implementation with a focus on the needs of particular local or Indigenous social groups and of the ecological situation'~\cite{Merchant1980}. The seven principles of data ecofeminism involve acknowledging the power relationships within climate change, recognising that the Global North is contributing more to it, considering the environmental costs of making AI by examining its materiality and supply chains, holding AI environmental costs accountable by publicly publishing environmental metrics, integrating a degrowth perspective, reclaiming digital sovereignty, protecting the digital and non-digital commons through mutual aid, and embracing Native epistemologies that have been dismissed in Western science and technology, and ultimately, aiming to build a world of many worlds where everyone fits.



\bibliographystyle{plain}
\bibliography{references}

\end{document}